# van der Waals epitaxy of lead monoxide – PbO and application as a virtual substrate for oxide membranes

*Marcel S. Claro*[1a], *Francisco Rivadulla*[1]

[1]*CiQUS, Centro Singular de Investigacion en Quimica Bioloxica e Materiais Moleculares, Departamento de Quimica-Fisica, Universidade de Santiago de Compostela, 15782 - Santiago de Compostela, Spain*

*a) Author to whom correspondence should be addressed:* marcel.santos@usc.es

ABSTRACT. We report the successful growth of epitaxial layers of van der Waals (vdW) lead oxide (PbO) polymorphs on different substrates, via pulsed laser deposition. A thin layer of vdW α-PbO (~5 nm) was subsequently used as a virtual substrate to grow thin films of perovskites $BaTiO_3$, $SrTiO_3$, and the spinel $CoFe_2O_4$ with a crystal quality comparable to direct epitaxy on single-crystal oxide substrates. Notably, films with the larger lattice parameter underwent spontaneous, controlled spalling within days of growth, remaining structurally intact, or were easily exfoliated and transferred using the tape method. This work establishes PbO as a promising virtual substrate technology for the fabrication of free-standing oxide membranes through vdW epitaxy, offering advantages over conventional sacrificial layer methods.

Lead oxides are recognized as key components in photovoltaic devices, transparent conducting films, gas and optical sensors, and X-ray imaging detectors due to their optical and electronic properties. The α-PbO polymorph exhibits a layered structure of tetragonal symmetry (space group P4/nmm) – Figure 1a, characterized by lattice constants a = 3.99 Å and c = 4.83 Å. In this structure, the layers are stacked along the [001] crystallographic direction, and the plane of Pb $5s^2$ lone pair electrons contributes to van der Waals (vdW) bonding. On the other hand, the band structure of α-PbO may be tuned by the number of vdW layers, leading to a quantum confinement effect [3,4] and an anomalously large bandgap and exciton binding energies of about 1 eV [3,4] at room temperature, when the systen approaches the single-layer limit[5,6]. Thus, these layered semiconductors present a tunable indirect or direct bandgap of 2.4 - 4.0 eV, in the visible region, which makes them ideal for optoelectronic applications.

Meanwhile, the β-PbO polymorph crystallizes in an orthorhombic phase (space group Pbcm, a = 5.89 Å, b = 5.49 Å, c = 4.75 Å), [1]with a crystal structure formed by zigzag chains stacked to create a layered structure similar to black phosphorus[2].

Polycrystalline PbO thin films have been grown by electron-beam evaporation, reactive DC magnetron sputtering, reactive ion beam sputter deposition, pulsed laser deposition (PLD) from an oxide target or a metallic target, and molecular beam epitaxy (MBE)[1,3,4,7]. However, epitaxial growth of phase-pure PbO is a rather challenging task, due to the number of phases in its phase diagram with similar or higher stability, which drive its decomposition into more stable $Pb+O_2$, $PbO_2$, $Pb_2O_3$, or mixtures of α- and β-PbO [8].

Previous works of vdW epitaxy of $InSe_x$, and $GaSe_x$ demonstrated that varying the stoichiometry during growth and the substrate temperature makes it possible to select a specific phase between the several phases and polymorphs of Indium Selenides[9] (InSe, β-In$_2$Se$_3$, γ-In2Se3 or ) and Gallium Selenides[10] (GaSe, Ga$_2$Se$_3$).

Similarly, we report the growth of epitaxial, phase-pure PbO polymorphs by PLD on different single-crystal oxide substrates. There is a narrow range of O$_2$ partial pressures, temperatures and epitaxial strain that allows the stabilization of high-quality, flat, vdW layers. We then used α-PbO to grow high-quality perovskite and spinel films on top, which can easily detach from the vdW and be transferred to a different substrate.

Depositing samples on (001) SrTiO$_3$ across a wide range of O$_2$ partial pressure and temperature, we obtained the phase diagram shown in Figure 1b. There is a very narrow growth window for pure-phase PbO, with the optimal conditions for growing pure epitaxial α-PbO over STO at 8 to 1.5mTorr and 550 ºC. A fine-tuning of the oxygen pressure during deposition can be made if the appearance of Pb droplets on the surface of the films or signatures of reduced PbO$_2$ are observed in post-fabrication AFM (S.I. 1) and X-ray diffraction analysis (Figure 1c), respectively.

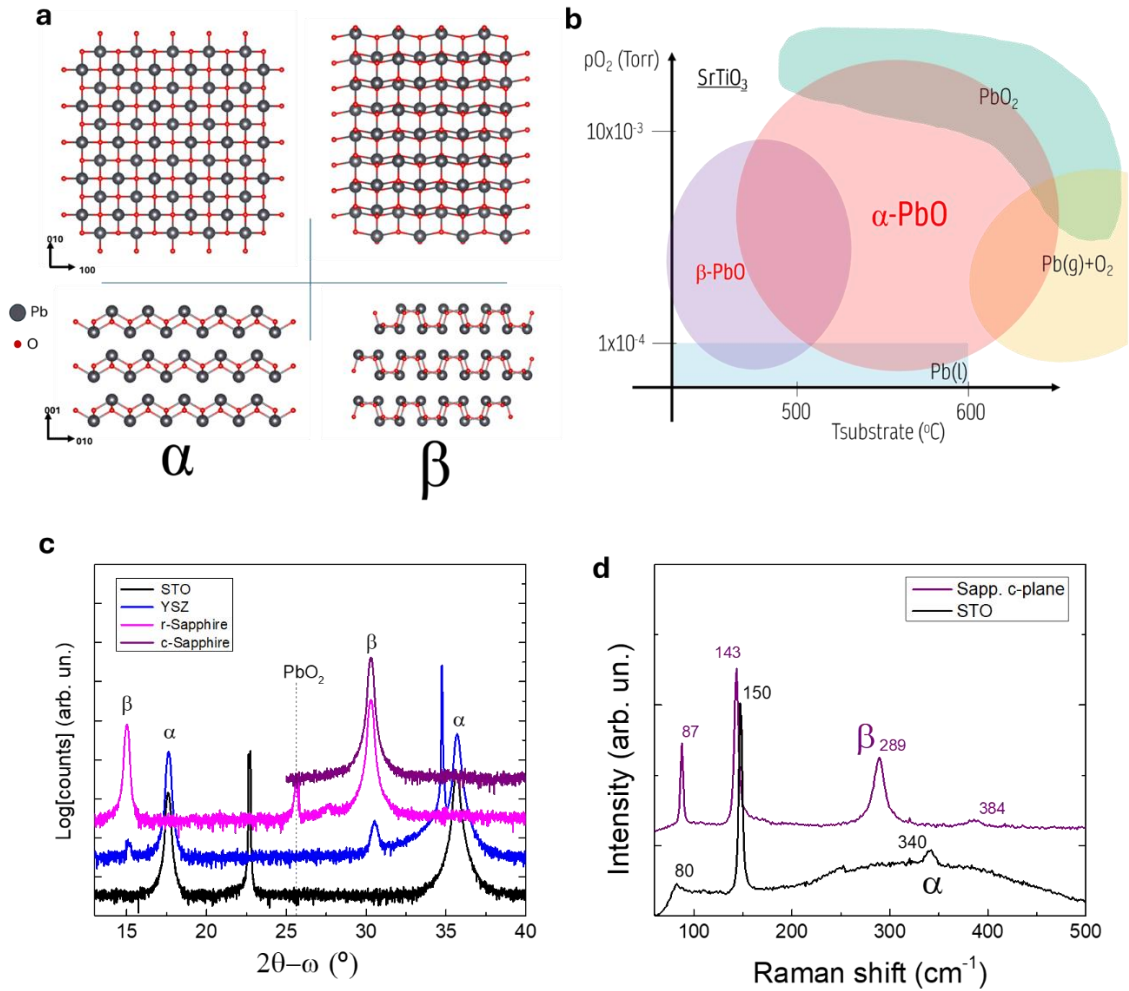

**Figure 1**. a) Lattice of PbO polymorphs b) Proposed phase diagram for growth of PbO on (001) SrTiO$_3$ substrates. There is a narrow stability window for the growth of the phase of each polymorph of PbO. As shown in the X-ray diffraction patterns and Raman spectra of panels c, d), it is possible to obtain pure α and β phases in STO and c-sapphire, or a mixture of both phases in other substrates.

The influence of epitaxial strain on the stabilization of the pure PbO polymorphs was studied by growing the PbO on different substrates: Y$_2$O$_3$:ZrO$_2$ (YSZ), STO, GdScO$_3$ (GSO), and c- and r-plane Al$_2$O$_3$ sapphire. While YSZ and r-sapphire resulted in the stabilization of a mixture of α and β phases of pure PbO, the phase pure α-PbO is stabilized on STO and GSO, while β-PbO grows on c-sapphire, which exerts the smaller compressive epitaxial strain on the films. This result reflects the importance of epitaxial stabilization energy in defining the interfacial reconstruction and the phase of the first layers. This result is very similar to $In_xSe_y$[9], where single-phase InSe is only obtained on a GaSe buffer layer, while the formation of mixed InSe and $γ-In_2Se_3$ phases is observed on *c*-sapphire or $α-In_2Se_3$, under the same conditions on graphene[11]. Note that for PbO the surface reconstruction is even more important than the chemistry of the surface itself. In fact, α-PbO grows on STO and GSO (Figure 2a and 4a), which have distinct chemistry but similar perovskite lattice; however, mixed-phases are obtained using c- and r-plane sapphire substrates. Recently, a lot of efforts have been made to fabricate standalone oxide perovskite membranes with various goals: recycling the expensive substrates; to explore the possibilities of controlling the strain in piezoelectric materials on flexible substrates or suspended membranes; or to transfer the functionality of epitaxial thin films to remote substrates. Up to now, most of the fruitful results come from vdW epitaxy over graphene[12,13] and sacrificial $SrAlO_x$ layers[14]. Layered PbO would have advantages over these methods: the vdW interface between the oxide film and the α-PbO allows the film to slide during the film release, contrasting to the sacrificial layer method, wherein strain is accumulated and only released suddenly during the etching of the sacrificial layer, which normally results in dislocation, defects, and cracks in the membranes. Another important advantage is that exfoliation could be done completely dry, without water or solvents that damage some of the perovskite films, e.g. $SrCoO_3$ and $SrFeO_3$, which become partially amorphous in water. Moreover, the growth of the PbO layer and the oxide film is done in two successive steps in a PLD chamber, which is a much simpler process than the graphene remote epitaxy.

There is, however, a possible drawback: the temperature for grow α-PbO and to preserve its long-term stability, which is lower than the usual temperatures for growing some oxides. Therefore, to use α-PbO as a virtual substrate it is necessary first to demonstrate that oxides can grow epitaxially over this layer and at temperatures low enough to maintain the structural and chemical stability of the thin layer of PbO. Thus, next we discuss the growth of crystalline perovskites and spinel's over a thin layer, between 5 and 7 nm, of α-PbO.

First, a heterostructure of 50 nm BaTiO$_3$/α-PbO @STO substrate was prepared at 550 ºC . The BaTiO$_3$ -BTO was chosen due to reported growth at temperatures as low as 310 ºC[15] and due to its technological interest as a ferroelectric and piezoelectric material. The lattice parameter of BTO is close to that of the PbO lattice (4.00 Å), accumulating strain over the STO substrate. In Figure 2a, the X-ray diffraction pattern of BTO/α-PbO confirms the pure phase α-PbO as well as crystalline (001) BTO signal; the strain imposed by the substrate produces a c-axis lattice parameter of BTO ~4.25 Å. The roughness of the surface is less than 1 nm RSM over 2 x 2 µm area, and is comparable to some epitaxial growth of BTO directly on STO[16]. Interestingly, we observed the full relaxation of the BTO layer during a few days after growth. The formation of the wrinkles was observed in real time, and the change of direction and branching could be observed for several seconds in the borders of relaxed areas. The XRD diffraction confirms that the BTO moves towards its relaxed lattice parameter (Figure 2a) after the process is complete on the whole substrate surface. Figure 2e shows a 50X optical microscopy image and an AFM image (Figure 2c) of the regular pattern formed after full relaxation of the oxide film. This intricate parameter is observed in BTO and other exfoliated piezoelectric membranes, and it is associated with the coupling between bi-axial strain and piezoelectricity during layer transfer[17–23]. Thus, these are indications that BTO is naturally spalling from the substrate, maintaining film properties, and paving the way for the development of a controlled and easy method for membrane extraction.

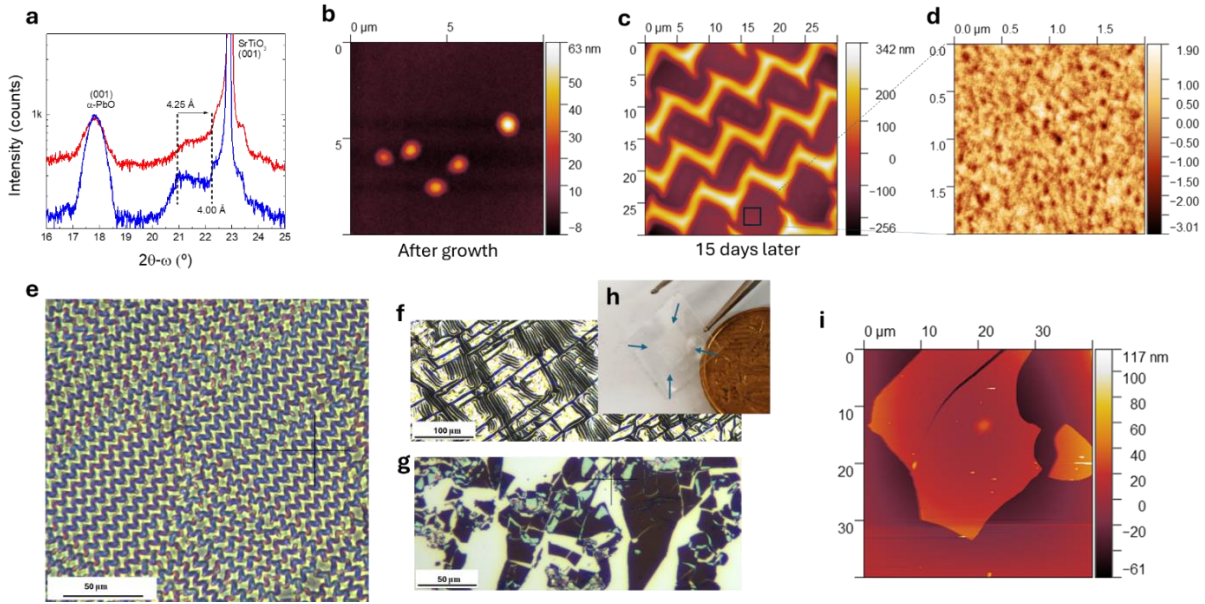

**Figure 2.** a) X-ray diffraction pattern of a) BaTiO$_3$ film on α-PbO@STO after growth (blue) and after relaxation, 15 days later (red). b) Atomic force microscopy (AFM) of BTO/α-PbO @STO a) after the growth and b) 15 days after the growth. c) Surface roughness in the flat areas in (b). Optical microscopy of the surface was performed more than 15 days after the growth, with the whole surface relaxed. e) exfoliated membrane in the adhesive tape and f) optical microscopy of the tape, g) flakes transferred from the tape to a Si (001) substrate, and g) AFM of flakes over the Si substrate.

The 50 nm BTO/7 nm α-PbO seen in Figure 2b-e was further characterized by transmission electron microscopy (TEM) with a lamella prepared in the [110] zone-axis. The BTO film lifts from the α-PbO layer at the wrinkles and remains attached to it in the flat areas between them. The EDS compositional maps (S.I. 2) indicate that the PbO vdW film remains mostly firmly attached to the substrate, with only some residues of PbO at the bottom of the suspended BTO membrane. Remarkably, the film's integrity is preserved in the whole lifting arc (Figure 3e, see also S.I. 4). The 2D Fourier transformer of selected areas shows the epitaxial relation between the BTO, PbO, and STO, with the spots [001] and [110] aligned. Nonetheless, the BTO film has secondary spots due to defects and polycrystallinity, and the measured lattice parameters point to relaxed BTO lattices.

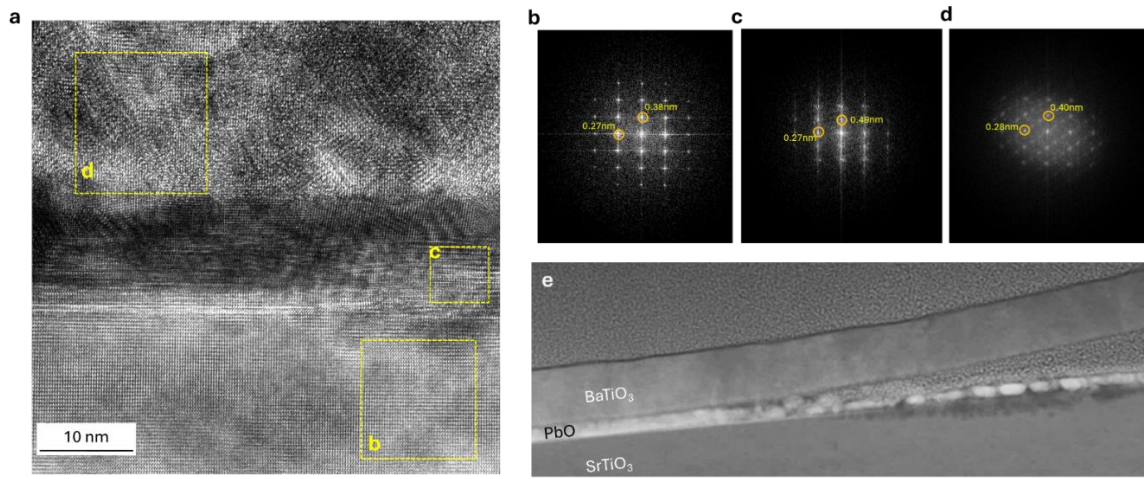

**Figure 3**. a) Transmission electron microscopy (TEM) of a BTO lamellae and Fast Fourier Transformer – FFT of selected areas of b) STO, c) PbO and d) BTO. Low magnification High-angle annular dark-field scanning transmission electron microscopy – HAADF-STEM close to one of the lifting points, showing a partially released BTO film from the PbO buffer layer, which remains attached to the substrate.

We have probed the exfoliation of the whole oxide film using the common adhesive-tape transfer method. The BTO/α-PbO@STO film just after the growth was easily exfoliated before the formation of the wrinkles, though some wrinkles were formed in the exfoliated membrane transferred to the tape (Figure 2f). Flakes of the BTO film of several microns wide were successfully transferred to Si substrates (Figure 2g). Consistent with the TEM observations, the XRD pattern of the remaining substrate shows the same PbO diffraction peaks as those observed in the original BTO/α-PbO@STO film, confirming that PbO remains attached to the substrate and is not lifted by the tape.

[24]BTO is often employed in memristive devices due to its ferroelectric properties, or as a tunneling barrier in ferroelectric tunnel junctions (FTJs)[25]. In the following, we demonstrate how our technique enables the rapid prototyping of heterostructures and devices based on these materials.

In Figure 2b, we show the X-ray diffraction pattern of a 20 nm STO/5 nm α-PbO heterostructure grown on GdScO$_3$ at 550 ºC. The GdScO$_3$ substrate was used in this case to observe the (00n) XRD peaks of the STO layer, which were separated from the substrate peaks. Additionally, the lattice parameter of the (100) GdScO$_3$ surface has a value of 3.97 Å, close to PbO parameter, which stabilizes the alpha phase of the oxide. As shown in Figure 4a), the (001) Bragg reflection of STO is clearly observed, as well as cubic periodicity in the $\phi$-scan (S.I. 5), and AFM confirms a roughness of around 1 nm (S.I. 6) in the flat areas, which resembles the homoepitaxy of STO/STO[26] at low temperatures. It makes STO ready to be used in conjunction with BTO. Nevertheless, in contrast to the BTO film, which forms wrinkles when spalling, the STO film shrinks: the calculated lattice parameter from XRD is the original 3.90 Å, creating cracks all over the surface (Figure 4b ), their size changes from sample to sample, indicating some relation with film thickness and cooling process.

Following, a multilayer stack consisting of 25 nm Nb:STO / 4 nm BTO / 25 nm Nb:STO was deposited on α-PbO@STO under conditions stated before. The structure was subsequently metallized by sputtering 2 nm Ti + 2 nm Pt together with a bare Nb:STO substrate. The stack was then transferred onto the metallized substrate using a thermal tape-assisted method. Figure 4c illustrates the schematic of the resulting heterostructure, while Figure 4d shows the morphology of the transferred flakes.

Electrical characterization was performed using micrometric probes in a probe station. The metallized Nb:STO substrate exhibited the typical memristive I–V characteristics of a metal/STO interface after electroforming and stabilization[24]. In contrast, the transferred flakes displayed a markedly different, more resistive behavior (Figure 7c), which dominates the overall transport properties. Both curves are stable after several cycles. This increased resistance can be attributed to electron blocking by the BTO barrier, combined with rectification effects at the Pt/STO interface. This experiment shows that it is possible to fabricate a multilayer functional device on top of vdW α-PbO, and do a dry transfer that retains its full functionality.

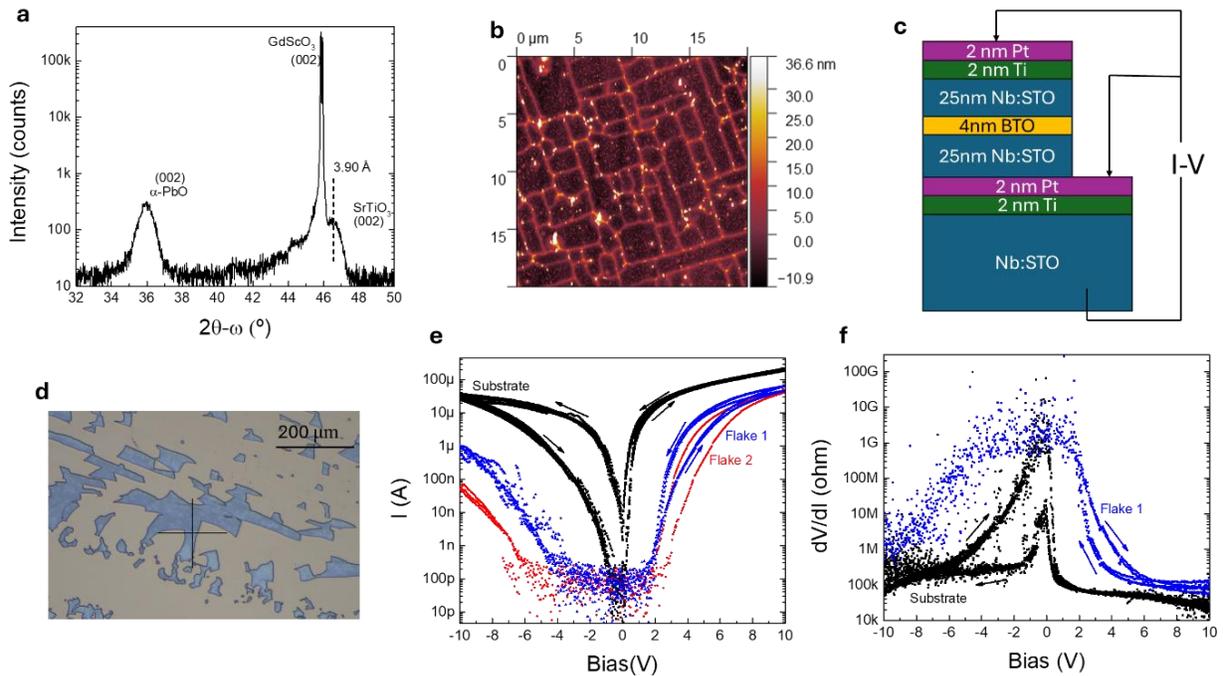

**Figure 4.** a) X-ray diffraction pattern of a 20nm film of STO grown on α-PbO@GdScO$_3$ just after growth. The vertical dashed lines indicate the lattice parameters of the oxide films, showing the partial relaxation. b) Atomic force microscopy (AFM) of 50 nm STO/α-PbO@GdScO$_3$ after the growth c) Schematic of material stack of the flakes over the substrate and the I-V measurement probe position. d) Optical microscopy of exfoliated flakes on top of metalized substrate. e) I-V hysteresis curve for the Nb:STO substrate and two different flakes. f) Differential resistance extracted from e).

As a representative case for AB$_2$O$_4$ spinel's, this method was applied also on CoFe$_2$O$_4$ - CFO. The CoFe$_2$O$_4$ is insulating with a large magnetic anisotropy. This, combined with a high Curie temperature, makes CFO thin films very appealing for the fabrication of spin-filter tunnel barriers. The grown CFO presented the same epitaxial quality, with single crystal diffraction (Figure 5a) and 0.4 nm surface roughness (Figure 5b) and could be easily exfoliated by the tape shown in the Figure 5c. While a 20 nm film did not show any wrinkle or crack after growth, a 50nm film was relaxed just after it. A relaxation pattern that somehow resembles a BTO pattern showed above (Figure 5f). Still, the pattern is less regular, as CFO has no piezoelasticity to regulate the wrinkle formation. Yet, similar zigzag-oriented branches are observed, suggesting this branching pattern is linked to the PbO piezoelasticity instead.

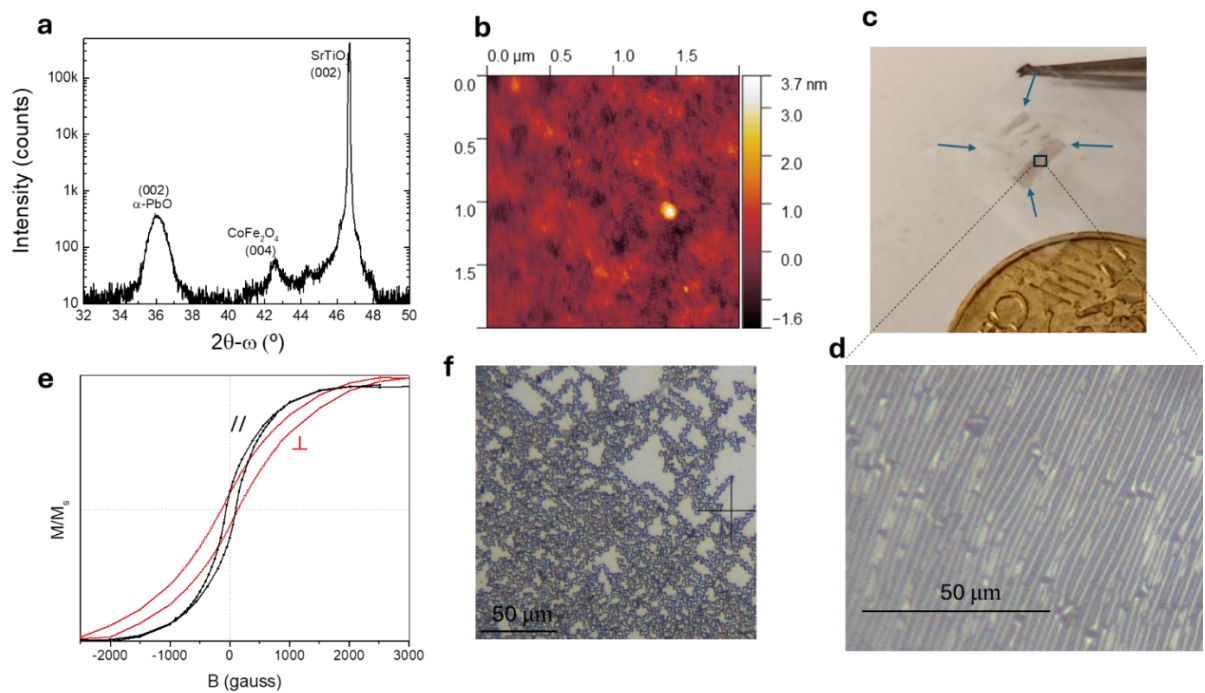

**Figure 5.** a) XRD pattern of CoFe$_2$O$_4$/α-PbO@STO and b) Atomic force microscopy image of the surface right after growth. c) Optical image of the exfoliated film on scotch tape, and d) enlargement of the image to show the continuous pattern on its surface. e) Magnetization of the CoFe$_2$O$_4$ membrane . f) Optical microscopy of the relaxed 50 nm CFO..

The magnetization of the 20 nm CFO membrane measured at room-temperature in Figure 5e shows a curve typical of a soft ferromagnet, with a reduced anisotropy probably due to redistribution of Co$^{2+}$ ions among the Oh and Th sites of the spinel by epitaxial strain.[27–31] Thus, in all cases, either perovskite or spinel, the material quality is similar to the epitaxy over single-crystal substrates, even if using growth temperatures lower than the usual. However, these layers tend to follow the current α-PbO surface lattice parameter and relax during the cool-down process or over a few days, spalling from the substrate. Depending on the strain, the film can shrink or wrinkle.

In summary, we have established growth conditions for pure-phase growth of α-PbO and β-PbO through the proper choice of substrate, temperature, and oxygen pressure. Subsequently, we used ~5 nm of α-PbO as a virtual substrate for the growth of perovskite and spinel oxides, presenting crystal quality similar to the epitaxy over single crystal substrates. A natural spalling of these perovskite membranes was observed, opening the way for new all dry methods for oxide membrane fabrication and transfer, which we demonstrated fabricating a memristive device based on BTO/STO heterostructure.

**Method**

**Pulsed Laser Deposition of the Films**

Thin films of PbO, $BaTiO_3$, $SrTiO_3$, and the spinel $CoFe_2O_4$ were grown by pulsed laser deposition from stoichiometric targets at an oxygen pressure of 2, 20, 100 and 10 mTorr respectively, with pulsed Nd:YAG $\lambda$ = 266 nm laser, 5 Hz and fluence of approximately 1, 1.5, 1.5 and 1.5 J cm$^{-2}$, respectively.

**Material Characterization**

Raman spectroscopy was measured at room temperature in a Witec alpha300 R confocal microscope, using a 50× objective lens, and a solid-sate 532 nm excitation laser. X-ray diffraction (XRD) measurements were performed in a PANalytical Xpert PRO MRD diffractometer with 5-axis cradle, standard Bragg–Brentano (BB) geometry, Cu anode X-ray tube operated at 45 kV accelerating voltage and 40 mA filament current to generate X-rays (Cu K-alpha). Soller and collimation 0.5″ slits were used in the source side, and a CCD detector (PiXcel), an inline (1D) model with additional Soller slit. Atomic force microscopy (AFM) measurements were taken under ambient air conditions with a Park NX20 in non-contact mode using *SSS*-NCHR cantilevers with a nominal tip radius of <5 nm, force constant of 42 N m$^{-1}$, and ≈350 kHz resonance frequency.

Cross-sectional TEM and STEM samples were prepared using a standard lift-out procedure in a dual-beam focused ion beam. Two protective Pt layers were deposited by electron beam and ion beam. The STEM and TEM images were acquired by a JEM-F200 Multi-purpose Electron Microscope operating at 200 kV. Field-dependent magnetization was recorded at room temperature in a Physical Property Measurement System (PPMS) from Quantum Design. Electrical measurements were performed by a Keithley 2450 source meter using tungsten probe tips.

ASSOCIATED CONTENT

Document with additional AFM, TEM, and XRD data.

**Author Contributions**

The manuscript was written through contributions of all authors.

**Funding Sources**

Any funds used to support the research of the manuscript should be placed here (per journal style).

ACKNOWLEDGMENT

F. R. acknowledges financial support from Ministerio de Ciencia (Spain), project PID2022-138883NB-I00, TED2021-130930B-I00, and Xunta de Galicia (Centro de investigación do Sistema Universitario de Galicia accreditation 2023–2027, ED431G 2023/03) and the European

Union (European Regional Development Fund – ERDF). The research of F.R. also received financial support from the Oportunius Program, Xunta de Galicia.

# van der Waals epitaxy of lead monoxide – PbO and application as a virtual substrate for oxide membranes

Marcel S. Claro[1a], Francisco Rivadulla[1]

[1]*CiQUS, Centro Singular de Investigacion en Quimica Bioloxica e Materiais Moleculares, Departamento de Quimica-Fisica, Universidade de Santiago de Compostela, 15782 - Santiago de Compostela, Spain*

a) *Author to whom correspondence should be addressed:* marcel.santos@usc.es

# Supplementary Information

S.I. 1 – Pb Droplets seen in the AFM (2 x 2 µm) and sample+tip photography

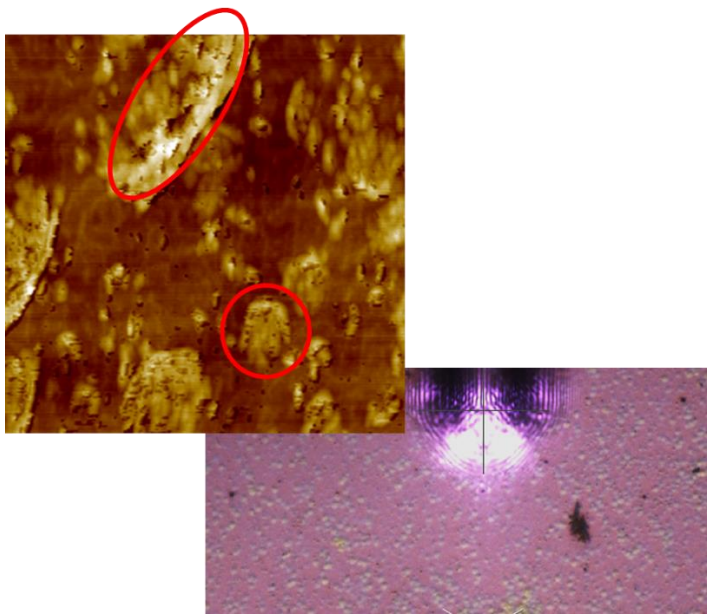

S.I. 2 – EDS of suspended BaTiO$_3$

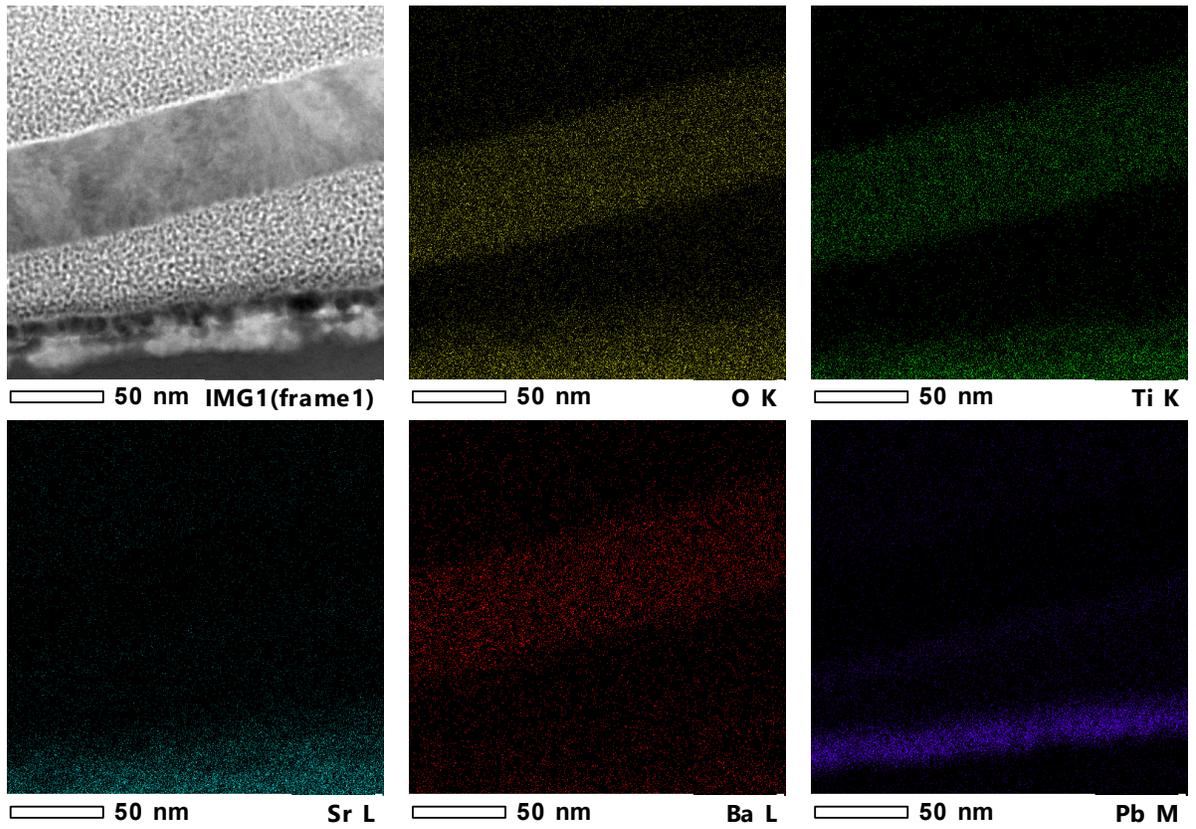

S.I. 3 - TEM and FFT of suspended BaTiO$_3$

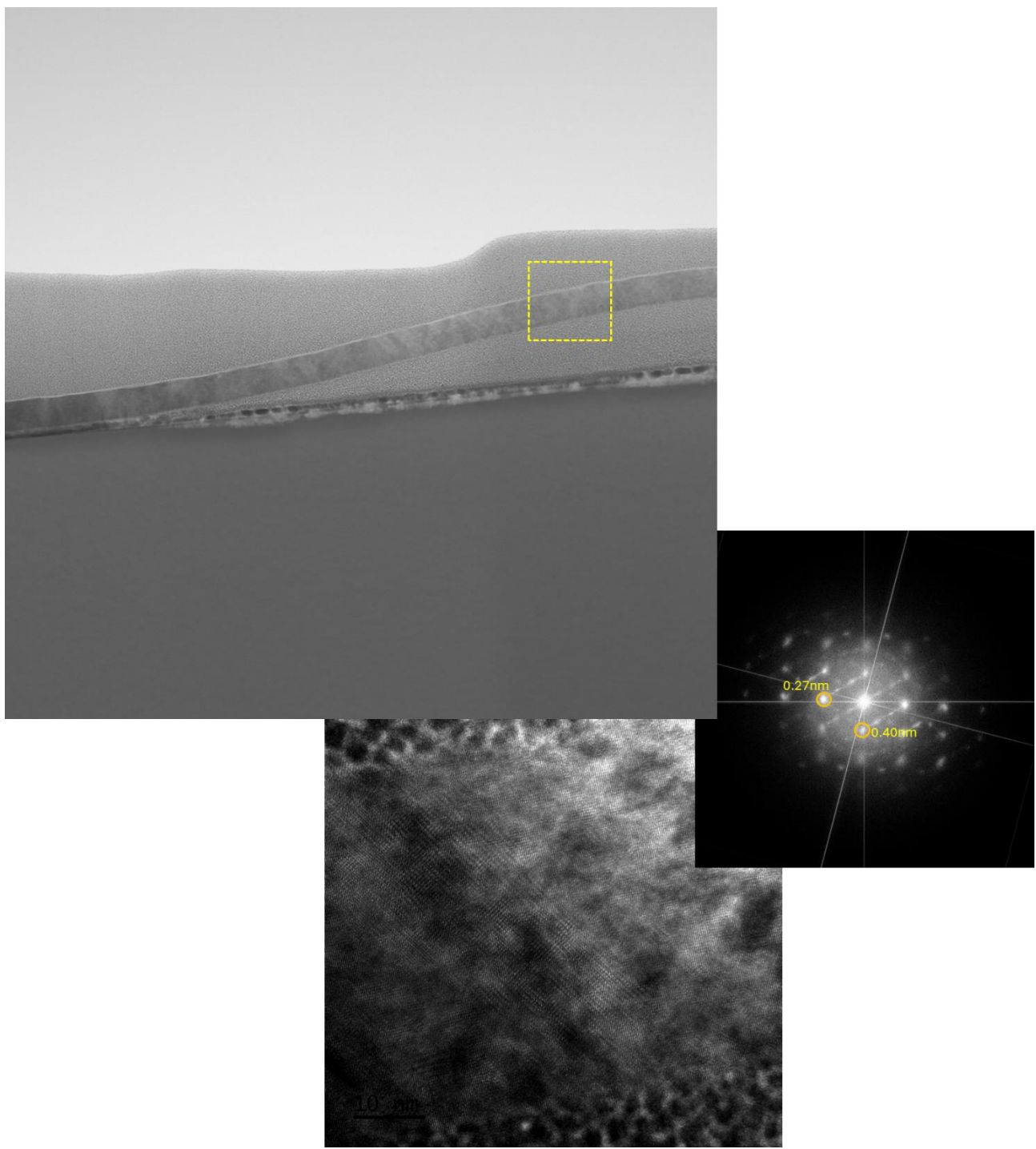

S.I 4 – φ-scan of [110] peak of 50 nm STO / 5nm PbO @ GdScO₃

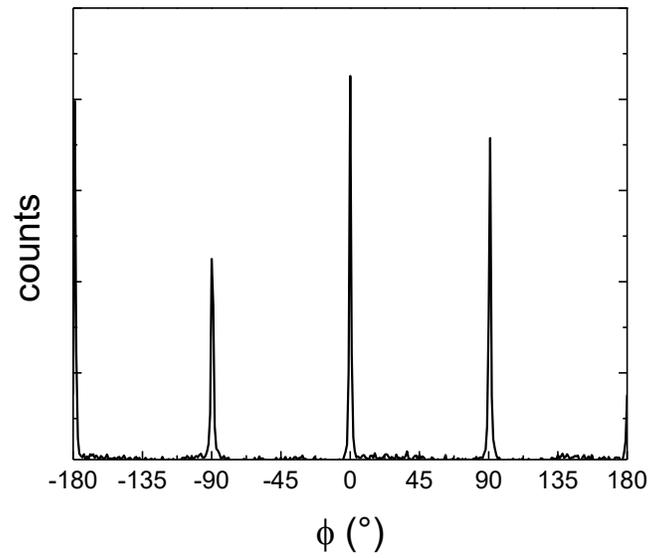

S.I 5 – AFM of 20 nm STO / 5nm PbO @ GdScO₃

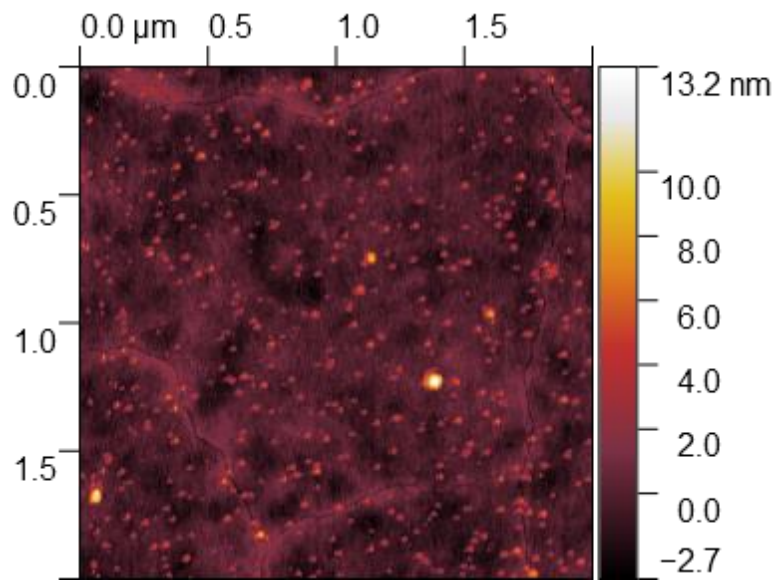